# Ambient Multimodality:
# an Asset for Developing Universal Access to the Information Society

*Noëlle Carbonell*

University Henri Poincaré, LORIA, CNRS & INRIA
LORIA, Campus Scientifique
BP 249
F54506 Vandœuvre-lès-Nancy Cedex
France
Noelle.Carbonell@loria.fr

## Abstract

Our aim is to point out the benefits that can be derived from research advances in the implementation of concepts such as ambient intelligence and ubiquitous/pervasive computing for promoting Universal Access to the Information Society, that is, for contributing to enable everybody, especially people with physical disabilities, to have easy access to all computing resources and information services that the coming worldwide Information Society will make available to the general public as well as to expert users in the near future.

Following definitions of basic concepts relating to multimodal interaction, the significant contribution of multimodality to developing Universal Access is briefly discussed. Then, a short state of the art in ambient intelligence research is presented, including references to some of the major recent and current research projects in this area. The last section is devoted to bringing out the potential contribution of advances in ambient intelligence research and technology to the improvement of computer access for physically disabled people, hence, to the implementation of Universal Access. This claim is mainly supported by the following observations:

(i) most projects aiming at implementing ambient intelligence focus their research efforts on the design of new interaction modalities and flexible/adaptive multimodal user interfaces; outcomes of these projects may contribute to improving computer access for users with physical disabilities;

(ii) these projects target applications meant to support users in a wide range of daily activities which will have to be performed simultaneously with the related supporting computing tasks; therefore, users will be placed in situational contexts where they will be confronted with similar difficulties to those encountered by physically disabled users, since they will have to share out their perceptual and motor capabilities between two tasks at least, a current daily activity and a computing task;

(iii) ambient intelligence applications being intended for the general public, a wide range of interaction devices together with flexible software for processing related modalities will be available, making it possible to provide physically disabled users with appropriate human-computer interaction tailored to their individual needs at reasonable expense.

## 1 Introduction

Ambient intelligence and ubiquitous or pervasive computing are concepts that refer to emerging technologies and potential software applications meant to populate our daily environment with smart artefacts and cooperative personal assistants. Numerous strategic objectives in the $6^{th}$ European Research Framework on Information Sciences and Technologies focus on issues relating to these concepts which motivate a growing interest from both researchers in human-computer interaction and user interfaces designers.

To implement these related concepts appropriately, a great variety of input and output modalities or combinations of modalities is necessary. In particular, adaptive multimodality is required for enabling efficient interaction with the "intelligent" artefacts and agents of all kinds which will soon assist us in our everyday life activities, hence in varied contexts of use: at home or in cars, on the street, on buses and trains, in public places such as shops, museums, etc.

Switching from one daily routine activity to another often implies switching interaction modalities. In addition, implementing ubiquity necessitates to accommodate user mobility, therefore to invent new interaction devices that can be embedded in ordinary objects belonging to the user's everyday environment, or in wearable objects. For instance, surfaces of everyday objects may be used as substitutes for pointing devices (Schmidt, Strohbach, van Laerhoven & Gellersen, 2003). Intelligent clothing or "interactive wear" may host a high diversity of computer applications (see, for instance, the Infineon web site). In addition, users' gestures, gaze, head and body movements can now be tracked and interpreted reliably, which makes it possible to further diversify the media and modalities offered to users for interacting with smart, local or remote, artefacts and agents.

While the diversification of context of use has favoured the development of new input modalities and combination of modalities, it has had but little influence on output media and modalities which are still much less diversified. Thanks to recent advances in speech synthesis, text to speech translation (Dutoit, 1996), interactive 3D visualisations (Ware, 2004), talking heads and embodied animated agents (Bosseler & Massaro, 2003), speech and graphics can be used effectively by designers for conveying a wide range of information, sensations and emotions to the user. However, further research is needed in order that haptics, sounds and odours may be considered as reliable easy to implement output modalities capable of enriching human-computer interaction.

Besides, thanks to the availability of reliable GPS information and wearable bio-sensors, useful information on the user's current environment and the interaction context can be inferred from the knowledge of their location, while being aware of their current physiological and emotional state may prove useful for assisting them. For instance, the context-aware MatchMaker prototype application, which was experimented on the CMU campus, can help a user to rapidly identify an expert user with appropriate knowledge for solving an unexpected problem or answering an urgent question. Awareness to spatial and temporal contexts enables MatchMaker to include proximity to the user and availability of the expert among the criteria used for selecting an expert (Smailagic et al., 2001).

Efforts to promote Universal Access in the coming Information Society could benefit from research advances in the implementation of ambient intelligence and pervasive computing. Such advances will enable the development of applications for the general public that will provide users with flexible multimodal interaction involving novel modalities with richer expressive power and greater transparency. Universal Access refers here to a long-term objective, difficult to attain: providing everybody with easy computer access.

At present, technological solutions exist that can satisfactorily compensate for many motor and perceptual handicaps. Numerous laboratory prototypes have been developed that implement them appropriately. However, only a few of them have been turned into commercial products, due mainly to the cost of developing and putting on the market appliances dedicated to small user communities. In addition, the diversity of motor and perceptual handicaps is such that most devices and software have to be tailored to individual capabilities and needs, which is impossible to achieve at reasonable expense, at least for the time being.

Besides, applications meant to implement the emerging concepts of ubiquitous computing and ambient intelligence, using novel interaction modalities and new user interface technologies, are fast developing. Such applications are likely to spread into the general public in the near future. Some possible side-effects of this evolution may prove beneficial and useful for promoting Universal Access. In particular, as the implementation of ambient intelligence implies the development of adaptive multimodal interaction based on a wide spectrum of media and modalities, this likely evolution will make it possible to propose a wide range of human-computer interaction devices and modalities to users with a great diversity of motor and/or perceptual disabilities. Designers and manufacturers will then have the means to offer them satisfactory computer access, namely, interaction facilities that can be appropriately tailored to their actual individual capabilities, and that can follow the possible evolution of their needs and requirements in the course of time. They will also have the opportunity to develop cheap flexible products from laboratory prototypes for compensating motor and perceptual handicaps appropriately. In addition, they will be able to propose novel assistance services aimed at helping users with physical disabilities to carry out everyday tasks, thanks to context awareness development tools (Dey, Salber & Abowd, 2001).

Current definitions of medium, modality, multimedia and multimodality will be presented in the next section together with a brief discussion of the potential contribution of multimodality to the implementation of Universal Access. Then, the state of the art in research on ambient intelligence, ubiquitous computing and "disappearing computer" will be briefly reviewed in section 3. Special emphasis will be laid on novel input and output devices, on new interaction modalities and combinations of modalities.

Section 4 will focus on proposing new research directions in assistive technologies that may benefit from recent scientific advances in the implementation of concepts such as ambient intelligence and ubiquitous computing, in the design of applications and user interfaces intended for the general public. We suggest research directions that may lead to fruitful applications, namely, applications contributing effectively to reduce computer access difficulties resulting from motor or perceptual handicaps, with a view to further extending the implementation of Universal Access design principles. In particular, a major aim of this reflection is to improve computer accessibility for users with physical disabilities that may evolve in the course of time, for instance as a corollary of age (cf. the growing perceptual and motor handicaps of senior users).

## 2 Definitions: medium vs modality, and multimedia vs multimodality

*Multimodality*, that is the simultaneous or alternate use of several modalities, appears as an emerging, useful interaction paradigm for advancing the implementation of universal accessibility.

*Modality* refers to the use of a medium, or channel of communication, as a means to express and convey information (Coutaz & Caelen, 1991; Maybury, 1993). Within the framework of a communication exchange, the sender translates concepts (symbolic information) into physical events which are conveyed to the recipient through an appropriate medium, and the recipient interprets the incoming signal in terms of abstract symbols. These processes involve the user's senses and motor skills and, symmetrically, the system input/output devices. According to the taxonomy presented in (Bernsen, 1994), several modalities or modes may be supported by the same medium (e.g., text and graphics used as output modalities). In other words:

> "… by *media* we mean the carrier of information such as text, graphics, audio, or video. Broadly, we include any necessary physical interactive device (*e.g.*, keyboard, mouse, microphone, speaker, screen). In contrast, by *mode* or *modality* we refer to the human senses (more generally agent senses) employed to process incoming information, *e.g.*, vision, audition, and haptics." (Maybury, 2001, p. 382)

Similarly, the terms *Multimedia* and *Multimodality* refer to two different realities. Multimedia systems have the capability of storing data conveyed through several input media, and to present them to the user through a set of appropriate output media. In addition, multimodal user interfaces are capable of resorting to several modalities for communicating information to the user (e.g., for generating system messages and responses); they have also the capacity to "understand" multimodal commands from the user, that is commands the information content of which is expressed using several modalities.

To characterize the various possible combinations of modalities or forms of multimodality, a taxonomy composed of four classes has been proposed (Coutaz & Caelen, 1991). As we focus here on issues relating to usage rather than implementation, we only need to consider two classes in this taxonomy:

- *alternate multimodality* [1], which characterizes a multimodal sequence of unimodal messages,
- *synergic multimodality*, which refers to multimodal utterances or, in other words, to the simultaneous use of several modalities in a single message.

The benefits that interface designers can draw from the implementation of multimodality are threefold. Firstly, alternate multimodality can advance computer accessibility, as it provides users with the means to choose among available modalities/media according to the specific usability constraints induced by the current context of use (Oviatt, 2000). Secondly, alternative media, modalities, and styles of multimodality are necessary for providing physically disabled users with computer access (Carbonell, 1999). Finally, synergic multimodality represents an indisputable improvement on unimodal interaction, in terms of expressive power, usability, and efficiency. For instance, the case study reported in (Cohen et al., 1998) suggests that multimodal interaction involving pen and speech may prove more efficient, in terms of speed mainly, than direct manipulation, at least for map-based tasks. According to the authors, "this advantage holds in spite of a 68% multimodal success rate, including the required error correction".

In particular, a significant advance in the implementation of universal computer access may be achieved, thanks to the integration of speech into standard user interfaces either as an *alternative* or a *supplementary* input modality. Integrating speech into user interfaces as a supplementary modality beneficially increases the limited expressive power of hand gestures and body movements. In addition, speech has major assets as a means of expression: it is flexible and currently viewed as the most natural human means of expression and communication. In addition, reliable speech processing technology is now available: the accuracy and robustness of present speech recognisers are sufficient to meet standard usability requirements. Products offering generic software tools for developing oral dialogue interfaces are available on the market (e.g., Speech SDK [2]). Appropriate software architectures and efficient "fusion" algorithms have also been proposed for integrating speech into multimodal user interfaces as an input modality (see, for instance, Nigay & Coutaz, 1993, or Johnston, 1998).

---

[1] This class results from the grouping of the "alternate" and "exclusive" multimodality classes in (Coutaz, and Caelen, 1991).

In the next section a tentative definition of the emerging concept of ambient intelligence is proposed. Then, some major scientific projects in this area are briefly described with a view to illustrating this definition.

## 3   Ambient intelligence and ubiquitous computing: an emerging research area

### 3.1   Definitions

*Ambient intelligence* refers to a broad scientific thematic including research activities on ubiquitous, pervasive and mobile computing mainly.
According to (Want, Farkas & Narayanaswani, 2005, p. 14):
> "*pervasive computing* aims to integrate computation into our daily work practice to enhance our activities without being noticed. In other words, computing becomes truly invisible."

Mark Weiser uses similar terms to characterise *ubiquitous computing* (see his web page):
> "it is invisible, everywhere computing that does not live on a personal device of any sort, but is in the woodwork everywhere."

Trends in computing and electronics are opening up new possibilities for interacting with computers. Faster networks and wireless technologies make it possible to interact with interconnected smart artefacts which behave in a coordinated way, using other devices than the standard peripherals of desktop PCs (i.e., mouse, keyboard, screen and speakers). These trends introduce new off-the-desktop application scenarios involving mobile users, mobile devices, and richer forms of interaction with cooperative artificial agents accessible from everywhere. The long-term objective of ubiquitous computing and ambient intelligence is to enable users to interact with a dynamic suite of devices embedded in their everyday environment so that they may be unaware of many of them. Phrases like "*disappearing computer*" or "*context-aware* systems/environments" refer to implications of these concepts.

In particular, *contextual awareness* is a key feature of ubiquitous computing systems. Context here means situational information relevant to the interaction between a user and an application:
> "A context-aware intelligent environment is a space in which a ubiquitous computing system has contextual awareness of its users and the ability to maintain consistent, coherent interaction across a number of heterogeneous smart devices." (Shafer, Brumitt & Cadiz, 2001, p. 363).

According to (Starner, 1999, p. 25-26), a context-aware interactive system must be capable of:
- perceiving the user and their interaction environment through various sensors some of which may be wearable;
- interpreting and modelling sensor data on the user (e.g., their current location, activity, physiological and emotional state), their physical environment, the interaction and task progress;
- interacting with the user through a contextually-driven user interface or distributed interaction devices embedded in everyday objects.

Ubiquitous computing and ambient intelligence imply mobile computing. Therefore, user mobility appears as a major design requirement for applications that aim at implementing these concepts appropriately. Obviously, monolithic standard GUIs meant for standalone PCs cannot meet this requirement. It is then necessary to break traditional styles of human-computer interaction. In particular, user interface functionalities have to be replicated in peripheral devices distributed in objects accessible to the user everywhere and at any time during their daily activities. These devices can be embedded either in objects in the user's everyday physical environment or in wearable objects, according to the activity the user is engaged in and the service or assistance they are meant to receive from the smart distributed system.
Interaction with such distributed user interfaces has to be flexible as it takes place in various physical settings and, most often, while users are engaged in some daily routine activity (e.g., driving, cooking). This specific context of use places constraints upon the selection of input and output modalities. These constraints which restrict the availability of the user's perceptual resources and limits the range of their physical actions need to be taken into account. As a side-effect, the volume and content of information exchanges between user and system have to be tailored to fit the specific expressive power and throughput of the chosen media and devices (e.g., PDAs, laptops, etc.). The implementation of these constraints raises new research issues which constitute a fast developing research area focused on achieving user interface *plasticity* (Calvary, Coutaz, Thévenin, 2001), especially through what may be called "adaptive multimodality".

---

[2] Speech SDK is a software toolkit developed by Philips for implementing speech command recognition and natural dialogue functionalities.

Additionally, the feasibility of user interfaces enabling interaction with smart artefacts embedded in everyday physical environments or in wearable objects offers the opportunity to design new styles of multimodal interaction. As for context awareness, the implementation of this concept, thanks to sensors capable of providing useful information on the user, the system and the interaction environment, makes it possible to extend the range of present computer applications, especially to design and implement novel, sophisticated forms of assistance to everyday life activities.

The next subsection presents a few research projects that contribute to the implementation of ubiquitous computing and ambient intelligence, especially by proposing realistic application scenarios or appropriate distributed computer environments and architectures, or new technologies for implementing context awareness and flexible multimodal interaction.

### 3.2  Examples

*EasyLiving* is one of the first major projects [3] that attempted to implement intelligent everyday environments (see the Microsoft web site). It includes a test laboratory equipped with distributed sensors and input-output devices embedded in the daily physical environment of the user. Main software tools include:
- for achieving context awareness by capturing situational information relevant to the interaction: computer vision (person tracking) and sensors (tracking of events, changes in the physical environment, etc.), geometric model of the physical environment;
- for providing flexible multimodal interaction: multiple sensor/actuator modalities, modality fusion, user interface adaptation;
- for facilitating the development of applications: automatic or semi-automatic calibration of sensors, and building of a geometric model of the user physical environment;
- for ensuring system extensibility: device-independent communication and data protocols.

In (Shafer et al., 2001), a realistic scenario of interaction with EasyLiving is described to illustrate the new interaction facilities that a context-aware intelligent environment can offer to users in their daily activities at home. Lights in a room can be remotely turned on using a speech command. Sitting in the living room on a sofa that faces the TV automatically activates the display of a menu with various options (e.g., "Watch a movie", "Play music" …), etc.

Modalities used for information exchanges between the user and the distributed system can change according to the situational context (alternate multimodality). For instance, a switch - instead of a speech command - may be used to turn on the lights in a room, so as to avoid disturbing a light sleeper in an adjoining room. Similarly, the system may resort to speech synthesis to convey an urgent warning to the user when asleep, instead of displaying it.

In (Smailagic et al., 2001), a design framework for context-aware applications is proposed. It is based on the assumption that "context aware applications are built upon at least two fundamental services, spatial awareness and temporal awareness" (p. 38), where temporal awareness amounts to the scheduled time of public and private events. As mobile computing is often performed simultaneously with another attention consuming primary task (e.g., walking, driving, etc.), the user may be distracted from their primary activity or neglect the other less important activity (i.e., computing). The aim of the Activity/Attention framework is to reduce user distraction and optimise the simultaneous execution of a primary task and a mobile computing task, using a Distraction matrix that characterizes activities by the amount of attention they require. Individual computer activities [4] are categorized by the amount of distraction they introduce in the primary activity.

---

[3] It started in 1997 and seems to have been active until 2001.
[4] Authors consider three classes of computer activities: information, communication and creation tasks.

Authors present some interactive applications that were developed to validate this design framework, demonstrate how to use it, and illustrate the benefits that can be gained from using it. For example, the Portable Help Desk (PHD) enables a mobile user to get maps of their immediate vicinity, including static and dynamic information. Thus, while walking on a campus, the user can obtain information on the location, contact information and availability of members of the academic staff, using their PHD.

More recent or current research projects are focused on further developing and implementing core concepts in pervasive computing and ambient intelligence.

The *Oxygen* project at MIT aims at supporting highly dynamic and varied human activities at home, at work and on the go. This ambitious project lays emphasis on enabling users to work together with other people through space and time (see the Oxygen web site).

*Ozone*, an EU funded project (5$^{th}$ FWP, 2001-2004), pursued roughly similar objectives, since it aimed to "investigate, define and implement/integrate a generic framework to enable consumer oriented ambient intelligence applications." The only difference lies in the user tasks and activities to be supported. While Oxygen was focused on supporting collective work, the main utility and usability objectives of Ozone were to "enhance the quality of life by offering relevant information and services to the individual, anywhere and at any time". A major outcome of this European project is the implementation of research results in two demonstrators, one for home activities, the "home demonstrator", and the other one for nomadic use, the "away demonstrator" (Ozone Final Report, section 5.5, 2005). See also (Diederiks, van de Sluis & van de Ven, 2003) for applications involving home activities.

The *CHIL* and *AMI* European projects (6$^{th}$ FWP) focus on enhancing and augmenting communication between users; see the web sites of these projects in progress.

CHIL (Computers in the Human Interaction Loop, 2004-2006) aims at creating "environments in which computers serve humans who focus on interacting with other humans". This goal will be achieved thanks to the design and implementation of computer services that will be capable, from the observation of people interacting face-to-face with each other, to infer the state of their activities and their intentions. These services will be endowed with perceptual context awareness to provide, whenever possible, implicit assistance which requires a minimum of human attention compared to explicit interaction. The main research issues addressed include:

- multimodal recognition and interpretation of all available perceptual cues in the environment with a view to explaining human collaborative activities and anticipating participants' intentions,
- quantifiable cognitive and social models of interactions between humans.

Smart rooms, namely office and lecture rooms, are the privileged applications that CHIL partners will use for demonstration and evaluation purposes.

AMI (Augmented Multi-party Interaction, 2004-2006) focuses on the design and implementation of new multimodal technologies capable of supporting human interaction in the context of smart meeting rooms and remote meeting assistance agents. It aims to make human interaction more effective in real time by developing new tools for supporting cooperative work and assisting navigation in multimodal meeting recordings. The project addresses several difficult research issues in two main areas:

- multimodal input processing, namely multilingual speech recognition and speaker tracking, gesture interpretation, handwriting recognition and shape tracking, integration of modalities and multimodal dialogue modelling;
- content abstraction, including multimodal information indexing, summarizing and retrieval.

Envisaged outcomes of the project include several demonstrators (e.g., an offline meeting browser and an online remote meeting assistant), and dissemination of large collections of annotated multimodal meeting recordings.

## 4    Ambient multimodal intelligence contribution to advancing Universal Access

As shown in the preceding section, past and present research on the implementation of concepts such as ubiquitous or mobile computing, pervasive or ambient intelligence has led to significant scientific and technological advances which have great potential to promote and stimulate the progress of research dedicated to the implementation of Universal Access (i.e., computer access for everybody) in the coming Information Society. In addition, they open up new possibilities for providing physically disabled people with enhanced assistance in their daily activities, thus improving their social integration.

All research attempts at implementing ambient intelligence (see section 3) aim at assisting users in their daily activities. In this context, computing tasks are viewed as secondary tasks that simply contribute to the completion of primary daily activities. They have then to be carried out in parallel with primary activities. This explains why most research projects on ambient intelligence lay emphasis on the design of new human-computer interaction modalities and flexible multimodal interaction so as to enable users to carry out computing tasks simultaneously with everyday activities without disrupting these activities. It also implies that users of these future ambient intelligence environments will be placed in situational contexts where they will be confronted with similar difficulties to those encountered by physically disabled users, since they will have to share out their perceptual and motor capabilities between two tasks at least, a current daily activity and a computing task.

Therefore, many solutions initially designed for enabling users of ambient intelligence applications to carry out computing tasks while performing some daily activity will prove appropriate for providing physically disabled users with easy computer access. Commercialised products meant for ambient intelligence users, such as new interaction devices with sophisticated modality and multimodality processing software, will also satisfy the needs and requirements of disabled users at reasonable expense, since ambient intelligence applications are meant for a very large user community including the general public. The great diversity of available new modalities will make it possible to provide physically disabled users with personalized and adaptable or adaptive user interfaces in order to take account of their specific disabilities which may evolve in the course of time [5]. Flexible synergic multimodality will be particularly useful for ensuring satisfactory computer access for users with complex disabilities (e.g., some kinds of palsies or visual deficiencies). Thus, significant advances in the implementation of Universal Access will be made rapidly and at reasonable cost.

For instance, successful implementation of alternate multimodality to ensure user interface plasticity may prove very useful for providing perceptual or motor disabled users with easy access to electronic information services. Solutions have already been proposed, and products will soon appear on the market for adapting information presentation and input formats to interaction devices endowed with varying information exchange capabilities and supporting varied media and modalities. The requirements and needs of mobile/nomadic users being roughly similar to those of a wide range of physically impaired users, applying these solutions to the latter category of users will significantly improve their access to computers.

In particular, the "translation" of information expressed in one medium into another medium, which is required for achieving user interface plasticity or for automating search through a multimodal database, is also necessary for providing users suffering from some specific perceptual disability with access to data stored in the medium corresponding to their deficiency. Some such translations/conversions are difficult to automate and prove costly to perform manually. For example, visual information has to be indexed to be accessible to blind users; such indexing is also required in the framework of the AMI project (see section 3) in order to optimise search in the large collections of multimodal meeting recordings that will be collected and analysed during the project. Efficient assistance to search through multimodal databases, a future application for the general public, implies the capacity to index visual data. In the same way, converting audio information into another modality (e.g., text) may prove useful for improving both multimodal data processing and computer access for deaf users. Thus, convergence between the needs and expectations of both the general public and users with some physical disability is an invaluable asset for speeding up both the design of solutions that will meet the requirements of the latter users and their implementation into appropriate commercial products.

Context awareness also may contribute to improving computer access for users with motor or perceptual impairments in-as-much as it offers the possibility to embed interaction devices in ordinary objects in the user everyday environment.

In addition, context awareness has a great potential for promoting social integration of users with physical disabilities, beyond improving their access to computing and electronic information resources. Its main contribution to helping these users overcome their motor or perceptual handicaps lies in the new possibilities it opens up for increasing their autonomy and their effective participation in collaborative and social activities.

For instance, awareness to the user location makes it possible to guide visually impaired people through places unknown to them. Context aware agents can be used for monitoring senior users in the course of their daily activities at home. Such agents can also provide useful assistance in nursing people who suffer from severe diseases, especially chronic diseases, and help motor disabled people to complete daily tasks, thus contributing to their autonomy.

---

[5] Taking account of the possible evolution of disabilities is crucial for an increasing number of physically disabled users including seniors.

Some research projects on ambient intelligence, such as Oxygen and AMI, lay emphasis on supporting mobile computing and remote meetings or, for the CHIL project, on augmenting face-to-face communication. Outcomes of these projects will prove useful for enabling users with motor or perceptual disabilities to participate in face-to-face and remote meetings.

Context awareness together with user interface distribution all over the user environment may significantly contribute to improving the daily life and social integration of people with physical disabilities. These contributions, however, will not be further detailed, since they are beyond the scope of the present paper which is focused on the potential contributions of ambient intelligence to advancing Universal Access in the Information Society.

## 4    Conclusion

This paper is an attempt at demonstrating and illustrating the potential benefits that can be derived from research advances in the implementation of ambient intelligence - a concept that subsumes ubiquitous (or pervasive) and mobile computing - for promoting Universal Access in the coming Information Society. Promoting Universal Access here means enabling everybody, especially people with physical disabilities, to have easy access to all computing resources and information services that will soon be made available to both the general public and expert users all over the world.

Basic concepts relating to multimodality and ambient intelligence are explained in section 2 and subsection 3.1, respectively.

Then, a short review of recent and current research projects on the design of ambient intelligence environments and the development of appropriate tools for implementing them (see subsection 3.2) shows that main efforts focus on the following issues:

- creating new interaction modalities that take advantage of recent technological advances, especially wireless connections which make it possible to distribute sensor and actuators in the user environment, hence to distribute and embed user interface components in ordinary objects;
- developing flexible adaptable multimodal interaction in order to implement user interface plasticity, a major requirement for the implementation of mobile computing;
- designing and developing applications meant to support users in a wide range of daily activities at home, at work and on the go, which implies taking advantage of system context awareness, that is, the system capability to "perceive" the user and their physical environment.

Our claim is that advances in research on ambient intelligence and the development of applications implementing this concept both for the general public and expert users can contribute significantly to the implementation and spreading of Universal Access. This claim is discussed in section 4 where it is supported by the main following argument.

As targeted ambient intelligence applications aim at assisting users in a wide range of daily activities, these activities will have to be performed simultaneously with appropriate supporting computing tasks. Therefore, users of such applications will be placed in situational contexts where they will be confronted with similar difficulties to those encountered by physically disabled users in-as-much as they will have to share out their perceptual and motor capabilities between two tasks at least, a current daily activity and a computing task. The convergence between the needs and requirements of small user communities made up of people suffering from similar disabilities, and those of a very large community including both the general public and expert users is a great opportunity for these small user communities. They will indirectly benefit from the outcomes of the research and development efforts that will focus on satisfying the needs and expectations of the larger user community. Thus, the likely spreading, in the near future, of ambient intelligence applications into the general public will significantly contribute to advancing the implementation of Universal Access and its spreading into society.